\documentclass[twocolumn,prd,showpacs,floatfix]{revtex4}
\usepackage{color}
\usepackage{graphics}
\usepackage{epsfig}
\usepackage{amsmath}
\usepackage{amssymb}
\usepackage{amsthm}
\usepackage{amsmath,alltt}
\usepackage{graphicx}
\usepackage[latin1]{inputenc}
\usepackage[T1]{fontenc}
\usepackage{color}

\newcommand{\slsh}[1]{{\not \! #1}}

\newcommand{\bea}{\begin{eqnarray}}
\newcommand{\eea}{\end{eqnarray}}

\newcommand{\be}{\begin{equation}}
\newcommand{\ee}{\end{equation}}
\newcommand{\nn}{\nonumber}

\begin{document}
\title{Chiral and Parity Symmetry Breaking for Planar Fermions:
Effects of a Heat Bath and Uniform External Magnetic Field}

\author{Alejandro Ayala$^1$, Adnan Bashir$^2$, Enif Guti\'errez$^2$,
Alfredo Raya$^2$ and Angel S\'anchez$^{2,3}$}
\affiliation{
$^1$Instituto de Ciencias Nucleares, Universidad
Nacional Aut\'onoma de M\'exico, Apartado Postal 70-543, M\'exico
Distrito Federal 04510, M\'exico.\\
$^2$Instituto de F\1sica y Matem\'aticas,
Universidad Michoacana de San Nicol\'as de Hidalgo, Edificio C-3, Ciudad Universitaria, Morelia, 
Michoac\'an 58040, M\'exico. \\
$^3$Department of Physics, University of Texas at El Paso, El Paso, TX 79968, USA.\\
}

\begin{abstract}

 We study chiral symmetry breaking for relativistic fermions, described by a parity violating
Lagrangian in 2+1-dimensions, in the presence of a heat bath and a uniform external magnetic
field. Working within their four-component formalism allows for the inclusion of both 
parity-even and -odd  mass terms. Therefore, we can define two types of fermion anti-fermion
condensates. For a given value of the magnetic field, there exist two different
critical temperatures which would render
one of these condensates identically zero, while the other would survive.
Our analysis is completely general: it requires
no particular simplifying hierarchy among the energy scales involved, namely, bare masses,
field strength and temperature.
However, we do reproduce some earlier results, obtained or anticipated in literature,  corresponding
to special kinematical regimes for the parity conserving case. Relating the chiral condensate to the
one-loop effective Lagrangian, we also obtain the magnetization  and the pair production rate for 
different fermion species in a uniform electric field through the replacement $B\to-iE$.

\end{abstract}

\pacs{12.20.-m,~11.30.Rd,~11.30.Er}

\maketitle

\date{\today}

\section{Introduction}

     Quantum electrodynamics in three dimensions (QED3) is of enormous and
recurrent interest due to its qualitative similarity with QCD. It exhibits
confinement and dynamical chiral symmetry breaking despite its tremendous
simplicity as compared to its non-abelian partner. Being super renormalizable, it
has no ultraviolet divergences, implying that its perturbative beta function
is zero. Therefore, it serves as an ideal framework within which
one can have a better understanding of the phenomena of confinement, dynamical
mass generation and the connection between them,~\cite{Bashir-Raya:2007-2009}.
Moreover, the infra-red behavior of a $d$-dimensional theory at high
temperature has been shown to be equivalent to the $(d-1)$-dimensional theory
at zero temperature,~\cite{Gross:1981}.
In particular, QED4 at high temperature has features equivalent to QED3 with coupling
$e^2 T$. Findings in~\cite{Pisarski:1981,Ayala-Bashir:2003} provide vivid examples
of this connection.

     There are many condensed matter systems whose low energy spectrum resembles that of
fermions in (2+1)-dimensions~\cite{Sharapov,supercon,ddw,lg,Tesanoviv:2001-2002}, including
graphene in the massless version~\cite{graphene}. In this connection, QED3 has been suggested to
strongly resemble the theory describing superconductor-insulator transition at $T=0$ and the
pseudo-gap phase in under-doped cuprates. There appears to exist a chiral symmetry 
at low enough energies in
standard $d$-wave superconductors. The destruction of the super-conducting phase
which leads to the appearance of the anti-ferromagnetism corresponds to the
spontaneous breaking of this chiral symmetry. This mechanism of spontaneous chiral
symmetry breaking has been seen to be formally analogous to the dynamical mass generation
in QED3~\cite{Tesanoviv:2001-2002}. Interest in the behavior of such systems in the presence
of a heat bath and external fields sparks a corresponding interest to explore QED3 under these
conditions.

     One should note that new features emerge in the underlying QED3 Lagrangian,
like the appearance of an additional mass term which is parity non-invariant and is
associated with a second fermion condensate in the theory. The Lagrangian thus describes
two fermion species which, in a convenient ``flavor''-basis, are non-degenerate
in mass and describe a light and a heavy fermion. Such parity violating Lagrangians
are relevant, for instance, in several four-fermion interaction models~\cite{cond4F}.
Moreover, in the context of the phase transition in $Bi_2Sr_2CaCu_2O_8$ observed by
Krishana {et. al.},~\cite{Krishana-1997}, it has been suggested~\cite{Laughlin-1998} that
parity and time reversal violating planar models could provide a plausible explanation
of the phenomenon observed at finite temperature and in the presence of a background
magnetic field, further triggering the need to study a model with these characteristics.

In this article, we study parity violating QED3 with a 4-dimensional reducible representation
of the Lagrangian. The breaking of chiral symmetry is caused by adding the
mass term $m\bar\psi\psi$. Though we also have the term  $m_o\bar\psi\tau\psi$, with 
$\tau=[\gamma^3,\gamma^5]/2$, often referred to as the Haldane mass term~\cite{Haldane}, it 
conserves chiral symmetry.
Chiral and parity symmetry breaking give rise to the parity conserving condensate
$\langle \bar{\psi} \psi \rangle$ (related to $m$) and the parity violating condensate
$\langle \bar{\psi} \tau \psi \rangle$
(related to $m_o$), respectively. We can define convenient linear combinations of these condensates which
separate the sectors of different fermion species, light and heavy. We denote them as
$\langle \bar{\psi} \psi\rangle_-$ and $\langle \bar{\psi} \psi {\rangle}_+$, respectively,
and derive their explicit expressions in the presence of a magnetic field and a heat bath without
resorting to any
particular hierarchy among the energy scales involved, i.e., the bare mass, the magnetic
field strength $\sqrt{eB}$, and the temperature $T$. The effects of the external magnetic field
and the thermal bath are found to be diametrically opposed for both of them. 
Earlier studies of dynamical chiral
symmetry breaking with these ingredients, considered separately, have already demonstrated that
magnetic fields support the formation of condensates while temperature tends to destroy them.
Here, for given values of $eB$ and the parameter $a=m_o/m$, which accounts for parity-violation effects, 
there exist two different values of the critical temperature
$T^c_+$ and $T^c_-$. $T^c_+$
corresponds to $\langle \bar{\psi} \psi \rangle_+ \rightarrow 0$ while
 $\langle \bar{\psi} \psi \rangle_- \neq 0$ and  $T^c_-$ ensures
$\langle \bar{\psi} \psi \rangle^- \rightarrow 0$, maintaining
 $\langle \bar{\psi} \psi \rangle_+ \neq 0$.
We also compute the magnetization and then the pair production rate for the two
fermion species in a uniform electric field through the replacement $B\to-iE$.
The light fermions are produced more copiously than the heavy ones. Moreover,
this effect is enhanced for intense electric fields.

The article is organized as follows: In Sect. II, we start out by describing the symmetries
of the Dirac Lagrangian in a plane, including both parity
conserving and violating fermion mass terms. Sect. III is devoted to calculating both the condensates
in the presence of a uniform magnetic field and a heat bath. In Sect. IV,
we relate the fermion condensate to the one-loop effective Lagrangian and obtain the expressions
for the magnetization and the pair production rate for the two species of fermions.
Conclusions are presented in Sect. V.

\section{Fermions in a plane}

Let us briefly review the model we consider in this article. For details,
see  Ref.~\cite{planarfemions}.
Working with an ordinary $4\times 4$ representation for the Dirac $\gamma^\mu$-matrices,
it is plain that only three of them are required to describe the dynamics on a plane. We
can choose them to be $\{\gamma^0,\ \gamma^1, \ \gamma^2\}$.
As $\gamma^3$ and $\gamma^5$ commute with these three matrices, the corresponding massless
Dirac Lagrangian is invariant under two chiral-like
transformations $\psi \to e^{i\alpha \gamma^3}\psi$ and $\psi \to e^{i\beta \gamma^5}\psi$.
In other words, it is invariant under a global $U(4)$ symmetry with generators $1$, $\gamma^3$,
$\gamma^5$ and $[\gamma^3, \gamma^5]$.  This symmetry is broken by an ordinary mass term
$m\bar\psi\psi$. Notice, however, that there exists the Haldane mass term~\cite{Haldane}, which is
invariant under the chiral-like transformations~:
$m_o \bar\psi \tau \psi$.
The term $m\bar\psi\psi$ is even under parity $\cal P$ and time reversal
$\cal T$ transformations, whereas, $m_o \bar\psi \tau \psi$ is not.  The corresponding free
Dirac Lagrangian in this case has the form
\be
{\cal L}= \bar\psi(i\slsh{\partial} -m-m_o\tau)\psi\;.\label{redlag}
\ee
There are many planar condensed matter models in which the low energy sector can be
written as this effective form of QED3, for which the physical origin of the masses
depends on the underlying system~\cite{Sharapov}. Examples are $d$-wave cuprate
superconductors~\cite{supercon, Tesanoviv:2001-2002}, $d$-density-wave states~\cite{ddw},
layered graphite~\cite{lg} and graphene in the massless version~\cite{graphene}.
We choose the Dirac matrices as
\be
\gamma^0=\left(\begin{array}{cc} \sigma^3 & \phantom{-}0 \\ 0 & -\sigma^3\end{array} \right), \quad
\gamma^k=\left(\begin{array}{cc} i\sigma^k & \phantom{m}0 \\ 0 & -i\sigma^k\end{array} \right),
\ee
for $k=1,2$ and
\be
\gamma^3=i\left(\begin{array}{cc} 0 & I \\ I & 0 \end{array} \right), \;
\gamma^5=-i\left(\begin{array}{cc} 0 & -I \\ I & \phantom{-}0 \end{array} \right),\;
\tau= \left(\begin{array}{cc} I & 0 \\ 0 & -I \end{array} \right).
\ee
Each of the mass terms is associated with a condensate; $m$ is related to the ordinary condensate,
whereas $m_o$ is related to $\langle\bar\psi\tau\psi\rangle$. As mentioned before, the antonym
properties of the mass terms suggest that the Lagrangian describes two fermion species, which in
graphene correspond to the different species in each triangular sublattice of the hexagonal lattice. 
This can be comprehended easily if we work with chiral-like eigenstates rather than with parity 
eigenstates. For this purpose, it is convenient to introduce the chiral-like projectors
\be
\chi_\pm = \frac{1}{2}(1\pm \tau)\;,
\ee
which verify,~\cite{Kondo:1996},
$\chi_\pm^2=\chi_\pm$, $\chi_+\chi_-=0$, $\chi_++\chi_-=I$, along with the trace properties
\bea
Tr[\chi_\pm]&=&2\;, \nn \\
Tr[\gamma^\mu\chi_\pm]&=&0 \;.
\eea
The ``right handed'' $\psi_+$ and ``left handed'' $\psi_-$ fermion fields are given by
$\psi_\pm=\chi_\pm \psi$. The $\chi_\pm$ project the upper and lower two component spinors (fermion species)
out of the four-component spinor $\psi$. The chiral-like decomposition of the free Dirac Lagrangian 
then becomes
\bea
{\cal L}= \bar\psi_+(i\slsh{\partial}-m_+)\psi_++\bar\psi_-(i\slsh{\partial}-m_-)\psi_-\;,\label{chirallag}
\eea
where $m_\pm = m \pm m_o$, and it obviously describes two species of fermions, each with a
different mass, $m_\pm$. This allows us to define a more convenient set of condensates
\bea
 \langle\bar{\psi} \psi\rangle_\pm &=&  \langle\bar{\psi} \psi\rangle \pm
 \langle\bar{\psi} \tau \psi\rangle
\;.
\eea
We evaluate these condensates in the next section by relating them to the fermion propagator and
employing Schwinger's proper time method.

\section{The Condensates}

In the section, we calculate the condensates  $\langle\bar{\psi} \psi\rangle_\pm$ in the presence of a
thermal bath and a  uniform magnetic field perpendicular to the plane of motion of the fermions.
In terms of the fermion propagator, $\langle\bar{\psi} \psi\rangle_{\pm}=-Tr [S_{\pm}(x,x)]$.
Following the notations and conventions of~\cite{us},
within the Schwinger's proper time framework~\cite{schwinger},
the fermion propagator can be expanded over the Landau levels~\cite{alanchodos} as
\bea
   S_{\pm}(k)= i \sum^\infty_{l=0}
           \frac{d_l(\frac{k_\perp^2}{eB})D +
           d'_l(\frac{k_\perp^2}{eB}) \bar D}{k^2_{0}-2
           l eB-m_{\pm}^2
           + i\epsilon} \; \chi_{\pm} + \frac{\slsh{k_{\perp}}}{k^2_\perp}  \; \chi_{\pm} \;,
\label{ferpropsum}
\eea
where
$d_l(\alpha)\equiv (-1)^l e^{-\alpha}
L^{-1}_l(2\alpha)$, $d'_l=\partial d_l/\partial \alpha$,
$D = (m_{\pm}+\slsh{k_{0}})+ \slsh{k_{\perp}} (m_{\pm}^2-k^2_{0})/
k^2_{\perp}$ and
$\bar D = -\gamma^5 \slsh{u}\gamma^3(m_{\pm} + \slsh{k_{0}})$. Here
$L_l^m(x)$ are the associated Laguerre polynomials and $u^\mu=(1, \vec{0})$. In case of a heat bath,
the 4-vector $u^{\mu}$ describes the plasma rest frame.
Hence, the condensate acquires the form
\bea
\langle\bar\psi\psi\rangle_{\pm}
     =-i2m_{\pm} \int \frac{d^3k}{(2\pi)^3} \sum_{l=0}^{\infty}
      \frac{(-1)^l e^{-\frac{k_\perp^2}{eB}}L_l^{-1}(\frac{2k_\perp^2}{eB})}
           {k_0^2-2leB-m_{\pm}^2+i\epsilon}\;.
\label{chiral21a}
\eea
We calculate the effect of a heat bath
on the condensates within the imaginary-time formulation
of thermal field theory~(see, for example,~\cite{kapusta}). In this formalism, the
integration over time component $k_0$ is replaced by a sum over
Matsubara frequencies according to the prescription
\bea
\hspace{-2mm}   \int \hspace{-2mm} \frac{d^3k}{(2\pi)^3}f(k)\rightarrow
   T\sum_n\int\frac{d^2k}{(2\pi)^2}f(\omega_n,{\bf k}) \;,
\label{defmatsu}
\eea
where $\omega_n=(2n+1)\pi T$ for fermions, with $n=0,\pm
1,\pm 2,\pm 3\ldots$, and $T$ is the temperature.
Thus at finite $T$ and $B$, the condensate is given by
\bea
\Delta\langle\bar{\psi} \psi\rangle_{\pm}&=&
     \frac{m_{\pm}eB}{2 \pi}
      \sum_{l=0}^{\infty}
      \frac{(2-\delta_{0l})
      \tilde{n}(\sqrt{2leB+m_{\pm}^2})}{\sqrt{2leB+m_{\pm}^2}}\nonumber \\
      &+&\Delta\langle\bar{\psi} \psi\rangle_{\pm}^B \;,
\label{chiral21Ta}
\eea
where
$\tilde{n}(x) = (e^{\frac{x}{T}}+1)^{-1}$ is the Fermi-Dirac distribution.
The notation used is $\Delta\langle\bar{\psi} \psi\rangle
= \langle\bar{\psi} \psi\rangle - \langle\bar{\psi} \psi\rangle_0$, where the
subscript $0$ stands for the value in pure vacuum, i.e., for $B,T=0$.
The presence of the superscript $B$ on the last term on the right 
highlights the fact that this quantity is evaluated in the presence of a magnetic 
field but at zero temperature. 
We can
rewrite
Eq.~(\ref{chiral21Ta}) as follows~:
\bea
\Delta\langle\bar{\psi} \psi\rangle_{\pm}
     &=& \frac{m_{\pm}eB}{4 \pi^{\frac{3}{2}}}
      \int_0^\infty\frac{ds}{s^\frac{1}{2}}
       e^{-sm_{\pm}^2} \left[1 - \Theta_4 \left(0, {e}^{- \frac{1}{4 T^2 s} } \right) 
\right]\nonumber \\
      &\times&\coth(eBs)+
      \Delta\langle\bar{\psi} \psi\rangle_\pm^B\;,
\label{condensate21Tebpoiss}
\eea
where $\Theta_4(u,x)$ is the fourth Jacobi-theta function.
\begin{figure}[t!] 
\vspace{0.4cm} {\centering \resizebox*{0.44\textwidth}
{0.18\textheight}{\includegraphics{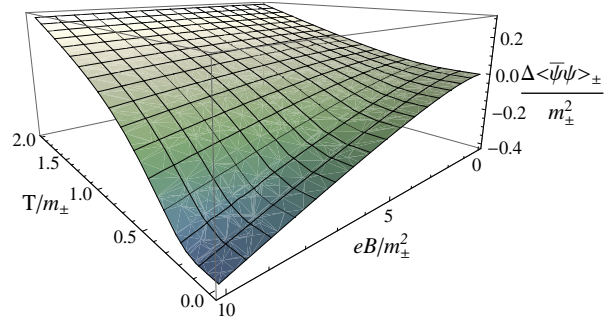}}
\par}
\caption{Temperature and magnetic field dependent fermion condensates as a function of
  $eB/m_{\pm}^2$ and $T/m_{\pm}$. }
\label{fig1}
\end{figure}
\begin{figure}[t!]
\vspace{0.4cm}
{\centering
\resizebox*{0.35\textwidth}
{0.3\textheight}{\includegraphics{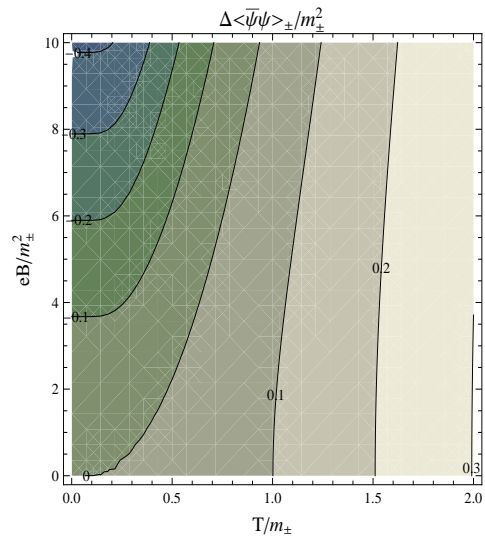}}
\par}
\caption{Contour plot for the temperature and magnetic field dependent fermion condensates 
as a function of
$eB/m_{\pm}^2$ and $T/m_{\pm}$.}
\label{fig2}
\end{figure}
The exact expression for the condensates is plotted in Fig.~\ref{fig1}. Due to the fact that 
the functional 
dependence of the condensates upon their respective bare masses  is the same, we have defined
dimensionless 
quantities in terms of the corresponding powers of $m_\pm$. The combined effects of the two agents on these 
quantities are such that at high temperatures and weak magnetic fields, the condensates are positive, but in 
the opposite regime when  temperatures are low and the  magnetic field is intense, it
pulls down the condensates to their large negative values. In the region where the condensates do not
deviate much from its zero value, temperature and magnetic field tend to nullify
the effect of each other, as can be better seen from  the contour plot displayed in Fig.~\ref{fig2}. Of course 
the numerical details which define the intense and weak magnetic field limit for each fermion species depend 
on the value of the parameter $a=m_o/m$.
As $\Delta\langle\bar{\psi} \psi\rangle_{\pm}=0$ would correspond to two different criticality curves, one
deduces that there exist two critical temperatures. $T_c^+$ renders
$\langle \bar{\psi} \psi \rangle_+ = 0$ while
 $\langle \bar{\psi} \psi \rangle_- \neq 0$ and  $T_c^-$ guarantees
$\langle \bar{\psi} \psi \rangle_- = 0$, preserving  $\langle \bar{\psi} \psi \rangle_+ \neq 0$.
This is better seen in Fig.~\ref{fig3}, where the behavior of $T_c$ as a function of $eB$  for different 
values of a mass parameter $\tilde m$ is displayed. 
The critical temperature behaves as
\be
T_c^2=\alpha({\tilde m}) {\tilde m}^2+\kappa({\tilde m})\left(eB\right)^{1-\delta({\tilde m})}
\label{fit}
\ee

\begin{table}
\begin{tabular}{|l|l|l|l|}
\hline
$\tilde{m}$ & $\alpha({\tilde m})$ & $\kappa(\tilde{m})$ & $\delta({\tilde m})$\\
\hline
1/20 & 0.0228 & 0.0583 & 0.0031  \\
1/15 & 0.0235 & 0.0584 & 0.0031 \\
1/10 & 0.0278 & 0.0586 & 0.0050 \\
1/5  & 0.0410 & 0.0586 & 0.0110 \\
1    & 0.0767 & 0.0627 & 0.0567 \\
5    & 0.3088 & 0.0807 & 0.1397 \\
10   & 0.8783 & 0.0977 & 0.1688 \\
15   & 1.4696 & 0.1079 & 0.2256  \\
20   & 2.1292 & 0.1144 & 0.2931 \\
\hline
\end{tabular}
\caption{Fit parameters for Eq.~(\ref{fit}) in units of ${\tilde m}_0=0.05$.}
\label{tabla1}
\end{table}

\noindent
The parameters are listed in Table~\ref{tabla1} for different multiples of ${\tilde m_0}=0.05$. 
Note that as $\tilde m\to0$, $\delta({\tilde m})\to 0$ and $\kappa({\tilde m})\to 0.0583$. 
For small values of mass, curves for the critical temperature are hardly 
distinguishable. A similar behavior was found in Ref.~\cite{Farakos:1998} for the 
parity-preserving case. Moreover,  the dependence of $T_c$ on mass becomes stronger 
as $\tilde m$ grows bigger. This implies that as $a\to 1$, $T_c^+$ depends more 
conspicuously on the mass,  whereas $T_c^-$ is roughly independent of it.

\begin{figure}[t!]
{\centering
\resizebox*{0.45\textwidth}
{0.23\textheight}{\includegraphics{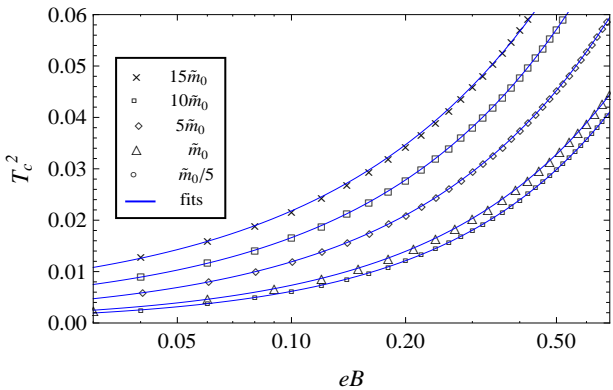}}
\par}
\caption{Critical temperature squared vs. magnetic field for different masses. Solid lines represent 
the fit in Eq.~(\ref{fit}).}
\label{fig3}
\end{figure}

\begin{figure}[t!] 
{\centering
\resizebox*{0.45\textwidth}
{0.23\textheight}{\includegraphics{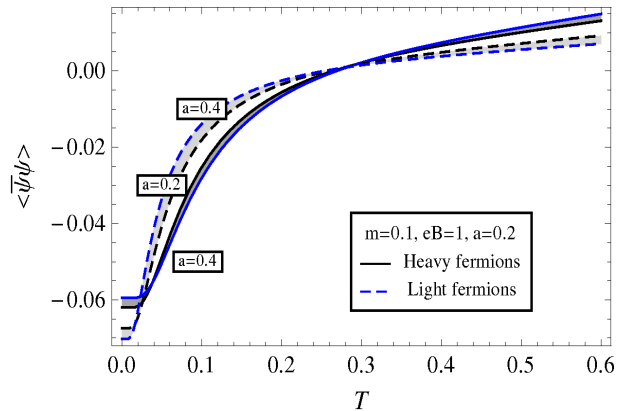}}
\par}
\caption{Temperature dependent condensates as a function of $T$ at fixed
$eB=1$ and $m=0.1$ for $0.2\le a\le 0.4$. {\it Light-shadowed region between dashed lines}: Light-fermion
condensate. {\it Dark-shadowed region between solid lines}: Heavy-fermion condensate.}
\label{fig4}
\end{figure}
Figure~\ref{fig4} displays the temperature dependent light- and heavy-fermion condensates  for
$0.1\le a\le 0.2$  at fixed $eB=1$ and $m=0.1$. We observe that for low temperatures, the effects of
the magnetic field are dominant, as expected. The lighter the fermions, the more visible these effects
are. As temperature increases, thermal effects start dominating the fermion condensation. In a selected
range  of values of $a$, we observe that the thermal effects for moderate values of $T$ are enhanced for light-fermions, but for
heavy-fermions, thermal effects are more visible at sufficiently higher temperatures.
\begin{figure}[t!] 
{\centering
\resizebox*{0.45\textwidth}
{0.24\textheight}{\includegraphics{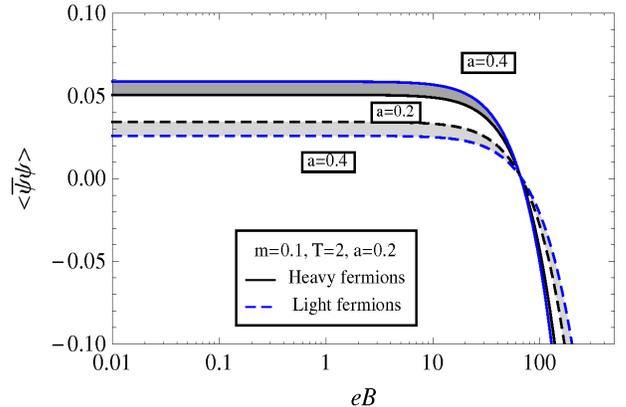}}
\par}
\caption{Magnetic field dependent condensates as a function of $eB$
at fixed $T=2$ and $m=0.1$ for
$0.2\le a\le 0.4$. {\it Light-shadowed region between dashed lines}: Light-fermion condensate.
{\it Dark-shadowed region between solid lines}:
Heavy-fermion condensate.}
\label{fig5}
\end{figure}

In Fig.~\ref{fig5}, we show the magnetic field dependence of the condensates for fixed values of 
$T$ ($T=2$) and $m$ ($m=0.1$). 
The parallel behavior of the two condensates seems to be affected only by the magnitude 
of the light- and heavy-fermion mass, and not necessarily by the parity-violating nature of $m_o$. Such an 
effect, however, might be evident at the non-perturbative level~\cite{Edward}.
In this connection, study of the dynamical mass generation and the corresponding calculation of the thermal 
conductivity is currently under way~\cite{Enif:2011}. Once the condensates have been evaluated, we can 
relate them with the Schwinger effect as we show in the next section.


\section{Effective Lagrangian}

An interesting application of our findings comes from the relation which exists between the
condensate and the one-loop effective Lagrangian ${\cal L}^{(1)}$~\cite{schwinger,gies}
\be
\langle\bar{\psi} \psi\rangle_{\pm} = - \frac {\partial {\cal L}^{(1)}_\pm}{\partial m_{\pm}} \;
\ee
in an external electromagnetic field. In this section we shall  obtain the magnetization of a gas of 
noninteracting fermions~\cite{andersen} through this effective Lagrangian in the presence of a constant 
magnetic field. As a second application, by replacing $eB\rightarrow -i eE$, we derive this quantity in 
an external uniform electric field and compare against the Schwinger formula for pair production.

\subsection{Magnetization}

Let us now proceed with the calculation of the magnetization of a gas of Dirac fermions of light and heavy 
species. The one-loop effective Lagrangian for each species at zero temperature in the presence of a 
magnetic field is simply given as
\be
{\cal L}^{(1)}_\pm = - \frac{1}{8\pi^\frac{3}{2}} \int_0^\infty \frac{ds\ e^{-sm^2_\pm}}{s^\frac{5}{2}}
\left[eBs \coth{(eBs)}-1 \right]\:.
\ee
Therefore, the magnetization 
$
M_\pm={\partial {\cal L}_\pm^{(1)}}/{\partial B}
$
for each species is
\bea
M_\pm
&=&- \frac{e}{8\pi^\frac{3}{2}} \int_0^\infty \frac{ds\ e^{-sm^2_\pm}}{s^\frac{3}{2}}\nn\\
&&\times \Bigg[\coth{(eBs)}-\frac{eBs}{\sinh^2{(eBs)}}\Bigg] \;. \label{MagvacallB}
\eea
In the weak field limit, expanding the expression in the square bracket and by direct integration, we find that
\be
M_\pm= - \frac{e^2B}{12 \pi m_\pm};\label{MagVac}
\ee
Expectedly, the magnetization grows linearly with the magnetic field and vanishes in its absence. 
Moreover, it is larger for the light species. 
The strong field case is better seen by writing Eq.~(\ref{MagvacallB}) in the equivalent form
\bea
M_\pm&=& - \frac{e m_\pm}{4 \pi }-\frac{e m^2_\pm }{ \sqrt{32} \pi  
\sqrt{eB}}\zeta \left(\frac{1}{2},\frac{m^2_\pm}{2
   eB}\right)\nn\\
   &&+\frac{3 e 
   \sqrt{eB} }{\sqrt{8} \pi }\zeta \left(-\frac{1}{2},\frac{m^2_\pm}{2
   eB}\right)\;, \label{Mstrong}
\eea
where $\zeta{(a,x)}$ is the Hurwitz-zeta function. Thus, taking the limit $m_\pm << \sqrt{eB}$, we find that
\be
M_\pm=\frac{3 e \sqrt{eB} \zeta   \left(-\frac{1}{2}\right)}{\sqrt{8} \pi }\;.\label{MagBst}
\ee
Here, $\zeta(x)$ is the Riemann-zeta function.
Thus the linear dependence gets transformed into a square-root dependence for intense magnetic fields.  
After taking into account the fcator of 2 for the fermion species, expressions~(\ref{Mstrong})
and~(\ref{MagBst}) compare correctly with the results which can be inferred from Eqs.~(A4,~7.4) 
of~\cite{Sharapov} respectively.

\begin{figure}[t!]
\vspace{0.4cm}
{\centering
\resizebox*{0.45\textwidth}
{0.24\textheight}{\includegraphics{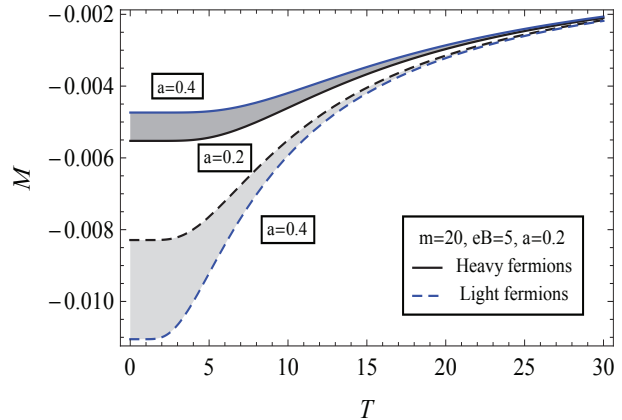}}
\par}
\caption{Magnetization as a function of $T$ at fixed $m=20$ and $B=5$ for $0.2\le a\le 0.4$. 
{\it Light-shadowed region between dashed lines}: Light-fermion. {\it Dark-shadowed region 
between solid lines}: Heavy-fermion.}
\label{fig6}
\end{figure}

Following the same reasoning, the effects of a thermal bath yield the following expression for the 
magnetization
\bea
M_\pm
&=&- \frac{e}{8\pi^\frac{3}{2}} \int_0^\infty \frac{ds\ e^{-sm^2_\pm}}{s^\frac{3}{2}} 
\Theta_4\left(0,e^{-\frac{1}{4 T^2 s}}\right)\nn\\
&&\times \Bigg[\coth{(eBs)}-\frac{eBs}{\sinh^2{(eBs)}}\Bigg]  \;. \label{magT}
\eea
Again, assuming $eB$ to be the smallest energy scale, we have that
\be
M_\pm=- \frac{e^2B}{12 \pi m_\pm}\left(1-2 \tilde{n}(m_\pm)\right)\;,
\ee
which reduces to  
\be
M_\pm=- \frac{e^2B}{24 \pi T} \;\label{magweak}
\ee
at high temperatures. 
Thus in the high temperature limit, we recover the free case, as can be observed in 
Fig.~\ref{fig6}. This is expected since in this regime, the effective Lagrangian approaches 
the thermodynamic potential for a gas of non interacting massive fermions. However, 
if $T\ll m_\pm$, the magnetization for each fermion species reduces to the vacuum case, Eq.~(\ref{MagVac}),
which agrees with the result of Ref.~\cite{andersen} for each species. 

\begin{figure}[t!] 
\vspace{0.4cm}
{\centering
\resizebox*{0.45\textwidth}
{0.24\textheight}{\includegraphics{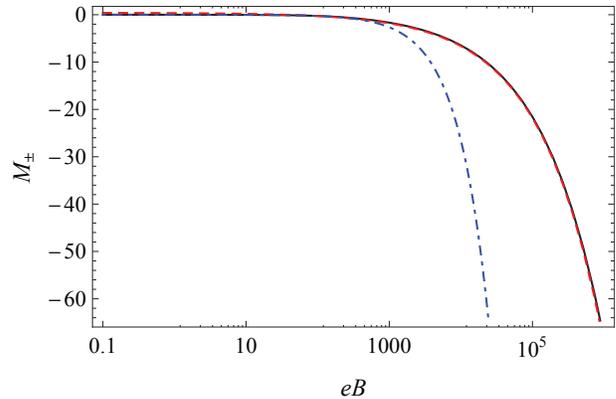}}
\par}
\caption{Magnetization as a function of $eB$ at fixed ${\tilde m}=0.05$ and $T=5$. {\it Solid-line}: 
Exact result, Eq.~(\ref{magT}); {\it Dashed-line}: Strong field limit, Eq.~(\ref{mstrongb}); 
{\it Dot--dashed-line}: 
Weak field result, Eq.~(\ref{magweak}).}
\label{fig7}
\end{figure}

In order to obtain the strong magnetic field behavior, let us focus on the thermal part of 
Eq.~(\ref{chiral21Ta}) alone. The vacuum part corresponds to Eq.~(\ref{MagBst}). We integrate term 
by term with respect to $m_\pm$ the infinite sum and then differentiate with respect to $B$. Thus 
the thermal part of the magnetization can be expressed as
\bea
M_\pm^T&=& - \frac{e}{2 \pi}\sum_{l=0}^\infty \left[\frac{eBl\  
{\tilde n}(\sqrt{2eBl+m_\pm^2})}{\sqrt{2eBl+m_\pm^2}}\right.\nn\\
&&\left. - T \ln (1 + e^{- \sqrt{2eBl+m_\pm^2}/T} )   \right]\;,
\eea
In the strong field limit, we have
\bea
M_\pm^T&=& - \frac{e}{2 \pi}\sum_{l=0}^\infty \left[\frac{\sqrt{eBl} \ {\tilde n}(\sqrt{2eBl})}{\sqrt{2}}\right.\nn\\
&&\left. 
- T \ln  \left(1 + e^{- \sqrt{2 e B l}/T } \right) \right]\;,
\eea
which is dominated by the $l=0$, magnetic field independent contribution. Higher Landau levels are 
exponentially suppressed. The contribution to the magnetization for the $l=0$ and $l=1$ levels, along with 
the vacuum contribution thus yields
\bea
M_\pm&=&-\frac{e \sqrt{eB} e^{-\frac{\sqrt{2eB}}{T}}}{\sqrt{8} \pi }+\frac{e T \ln
   (2)}{2 \pi }\nn\\
   &&+\frac{3 e \sqrt{eB}
   }{\sqrt{8} \pi }\zeta \left(-\frac{1}{2}\right)\;.\label{mstrongb}
\eea
The last term, which corresponds to the vacuum contribution, leads the behavior of the magnetization in 
this regime. The weak- and strong magnetic field behavior of the magnetization are illustrated in 
Fig.~\ref{fig7}. The strong field approximation,~Eq.~(\ref{mstrongb}), which is a much simpler expression
than~Eq.~(\ref{magT}), lies on top of the exact result for intense fields. When the fields are weak, 
the corresponding limit,~Eq.~(\ref{magweak}) describes the exact result perfectly well. Note that
it is not possible to appreciate the difference between the three results in that region due to the smallness
of $M_{\pm}$ and the fact that the plot has been drawn on a logarithmic scale along the $x$-axis.
Next we consider the pair production rate.

\subsection{Pair production}

\begin{figure}[t!]
\vspace{0.4cm}
{\centering
\resizebox*{0.45\textwidth}
{0.24\textheight}{\includegraphics{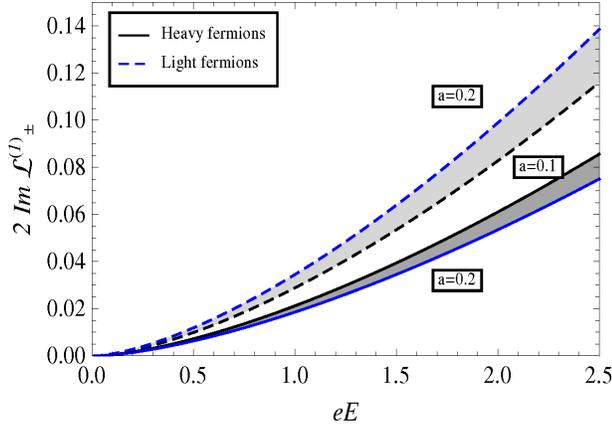}}
\par}
\caption{Pair production rate as a function of $eE$ at fixed $m=0.1$ for $0.1\le a\le 0.2$. 
{\it Light-shadowed region between dashed lines}: Light-fermion. {\it Dark-shadowed region between 
solid lines}: Heavy-fermion.}
\label{fig8}
\end{figure}

We can perform a consistency check of our findings against the Schwinger's formula for pair production,
Ref.~\cite{schwinger}. Let us remember that the Schwinger mechanism is understood in terms of an
imaginary part developed by the effective action due to the instability caused by an intense uniform
electric field. Such an imaginary part is readily obtained from the condensate by
simply replacing $eB\rightarrow -i eE$, i.e.,
\bea
\hspace{-2mm}     \Delta\langle\bar{\psi} \psi\rangle^E_{\pm}=
     -\frac{m_{\pm}}{4\pi^{\frac{3}{2}}}
     \int_0^\infty \frac{d\eta}{\eta^{\frac{3}{2}}}
     e^{-\eta m_{\pm}^2}\left[eE\eta\cot(eE\eta)-1\right].
\label{pair1}
\eea
Using the identity
\be
   \cot(eE\eta)=\frac{1}{eE\eta}+\sum_{n=1}^\infty\sum_{s=\pm1}
      \frac{1}{eE\eta+sn\pi}
\label{pair2}
\ee
in Eq.~(\ref{pair1}), we get
\bea
\Delta\langle\bar{\psi} \psi\rangle^E_{\pm}=
    -\frac{m_{\pm} eE}{4\pi^{\frac{3}{2}}}\sum_{n=1}^\infty\sum_{s=\pm1}
    \int_0^\infty \frac{d\eta}{\eta^{\frac{1}{2}}}
     \frac{e^{-m_{\pm}^2\eta}}{eE\eta+sn\pi}\;.
\label{pair3}
\eea
In order to obtain the pair production probability, we use the
prescription $m_{\pm}^2\rightarrow m_{\pm}^2+i\epsilon$ in Eq.~(\ref{pair3}).
It is equivalent to moving the cotangent poles by an amount ``$i\epsilon$''
above the $Re(\eta)$ axis. Thus, it is easy to show that the
imaginary part of the condensate in the presence of an electric field
is
\bea
Im[\Delta\langle\bar{\psi} \psi\rangle_\pm^E]&=&
   -\frac{m_{\pm}eE}{4 \pi^{\frac{1}{2}}}\sum_{n=1}^\infty\sum_{s=\pm1}
    \int_0^\infty \frac{d\eta}{\eta^{\frac{1}{2}}}
     e^{-m_{\pm}^2\eta} \nonumber \\
    &&\hspace{1cm} \times\delta(eE\eta+sn\pi)\nonumber \\
    &=& -\frac{m_{\pm}(eE)^{\frac{1}{2}}}{4\pi}
       \sum_{n=1}^\infty
       \frac{e^{-\frac{m_{\pm}^2}{eE}n\pi}}{n^{\frac{1}{2}}}\;.
\label{pair4}
\eea
Therefore, the pair production rate is,~\cite{Gusynin:1999},
\bea
2Im[{\cal L}^{(1)}_\pm]=\frac{1}{4\pi^2} (eE)^\frac{3}{2}
\sum_{n=1}^\infty \frac{e^{-(\pi m_{\pm}^2/eE)n}}{n^\frac{3}{2}}\;.
\eea
Recall that the total rate is given by the first term in the summation alone.
This result can be straightforwardly generalized to the case of parity violating mass terms
considering now the light- and heavy-fermion species, shown  in Fig.~\ref{fig8}.
Expectedly, increasing the intensity of the electric field, lighter fermions are much easier to
produce than heavy ones. Thus, parity-violating effects favor the production of the light species
over the heavy one.

\section{Conclusions}

We study the temperature and magnetic field dependence of the fermion anti-fermion
condensate for planar QED, allowing for the parity violating mass terms.
The 4-component study of this Lagrangian naturally leads us to consider two species
of fermions which are non-degenerate in mass. Correspondingly, there are
two types of condensates, with the magnetic field and temperature pulling them apart
in diametrically opposed directions. The effects of the external ingredients
are more pronounced for the lighter species of fermions as compared to the heavier one.
We carry out a detailed quantitative analysis of this statement. Moreover, we also
compute the magnetization induced by the magnetic fields and the pair production rate for 
both the species in the presence of an external
electric field. For the former case, the magnetization becomes independent of the mass of 
each species as the field increases, but it vanishes when the temperature gets higher. 
In the later case, with the increasing electric field intensity, it is relatively
easier to produce pairs of light species. For the mass $m=0.1$, the
production rate for the light pairs is twice as much as that of the heavy species for
the electric field intensity of the order of 3, as shown in the Fig.~\ref{fig8}. Keeping 
in mind that the parity violating
2+1-dimensional Lagrangian has a host of applications in various condensed matter systems,
we expect our results to have important bearing on their studies.

\section*{Acknowledgments}

We are thankful to V.P. Gusynin for his comments on the draft version of this article.
A.B and A.R acknowledge CIC (UMSNH), SNI and CONACyT grants. A.S and E.G are grateful to CONACyT for
a postdoctoral and a doctoral fellowship respectively.

\section*{APPENDIX}

Below we analyze different scenarios of relative strengths of the mass $m$ (for a general discussion
in this appendix, we adopt the notation ${\tilde m}$ instead of $m_{\pm}$
and $\langle\bar{\psi} \psi\rangle$ instead of $\langle\bar{\psi} \psi\rangle_\pm$
), the magnetic field $eB$
and the temperature $T$, namely, intense magnetic fields, {\em i.e.}, {$  {\tilde m} \ll T \ll \sqrt{eB}$},
intermediate magnetic fields, {$  {\tilde m} \ll \sqrt{eB} \ll T$}, and  weak magnetic fields,
{$ \sqrt{eB} \ll {\tilde m} \ll T$}. In these limits, we obtain uncomplicated and closed expressions.
What is interesting is that in some cases, these simple results can be used
instead of the complete Eq.~(\ref{condensate21Tebpoiss})  in all ranges of
relative strengths of the energy scales involved  without losing the quantitative precision
by a significant margin. Some of these results can be compared and contrasted with the ones
obtained in~\cite{Anguiano:2007,Farakos:1998}.
\begin{figure}[t!]
{\centering
\resizebox*{0.45\textwidth}
{0.25\textheight}{\includegraphics{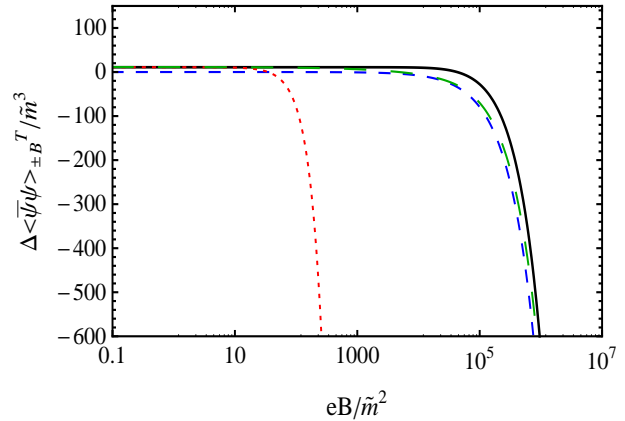}}
\par}
\caption
{Temperature and field dependent fermion condensate in the units of ${\tilde m}^2$
as a function of  $eB/{\tilde m}^2$ at fixed $T$. {\it Solid line}: exact result,
Eq.~(\ref{condensate21Tebpoiss}).
   {\it Dashed line}: strong field limit, Eq.~(\ref{chiral21Te}). {\it Long-dashed
     line:} intermediate field limit, Eq.~(\ref{chiralintereb}). {\it Dotted line}:
weak field limit, Eq.~(\ref{chiral21Tf}).}
\label{fig9}
\end{figure}
\begin{figure}[t!]
{\centering
\resizebox*{0.45\textwidth}
{0.25\textheight}{\includegraphics{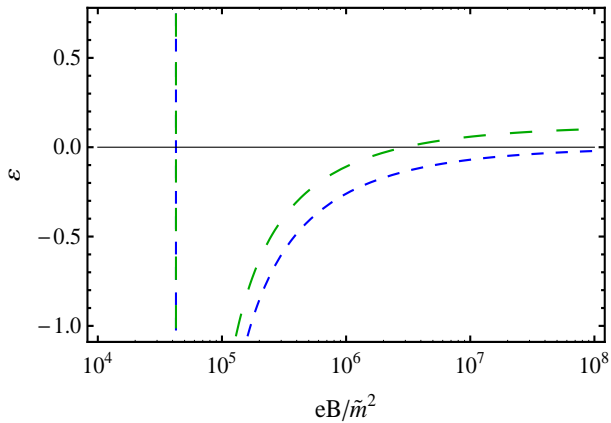}}
\par}
\caption{Relative error of the condensate. {\it Dashed line}: strong field asymptotics,
Eq.~(\ref{chiralintered}). {\it Long-Dashed line}: intermediate field asymptotics, Eq.~(\ref{chiralinterec}).}
\label{fig10}
\end{figure}

\subsection{Strong field limit}

Assuming the strength of the magnetic field to be the largest of the energy scales involved, the main
contribution to the condensate comes from the lowest Landau level, $l=0$, which  takes the form
\bea
      \Delta\langle\bar{\psi} \psi\rangle_B^T=
     -\frac{eB}{4\pi}
       \left[
       1-2\tilde{n}({\tilde m})
       \right]\;.
\label{chiral21Te}
\eea
Furthermore, when ${\tilde m}\ll T$, the leading
contribution to the condensate is then given by
\bea
 \Delta\langle\bar{\psi} \psi\rangle_B^T=
     -\frac{eB}{8\pi}\frac{{\tilde m}}{T}\;.
\label{chiralintered}
\eea
A comparison against the exact result is shown in Fig.~\ref{fig9}. The dashed curve, that corresponds
to the expression in Eq.~(\ref{chiral21Te}), approaches the exact result (solid curve) as the strength
of the field increases.

\subsection{Intermediate Field }

In order to obtain the intermediate field limit, we perform a high temperature expansion in the Fermi-Dirac
distribution  in Eq.~(\ref{chiral21Ta}). The leading term in this series can be written
as follows
\bea
      \Delta\langle\bar{\psi} \psi\rangle_B^T
    &=&\frac{{\tilde m}eB}{4 \pi^{\frac{3}{2}}}
       \int_{\frac{3}{2\pi^2T^2}}^\infty\frac{d\eta}{\eta^\frac{1}{2}}
       e^{-\eta {\tilde m}^2}\coth(eB\eta)\nn\\
       &&-\frac{eB}{4 \pi}\;,
       \label{chiralintereb}
\eea
where we have introduced an ultraviolet cutoff $(2/3)\pi^2T^2$ due to the
fact that in the limit $eB\rightarrow 0$, each component of the
transverse momentum contributes to the thermal bath with a factor
$(1/3)\pi^2T^2$. Notice that in Fig.~\ref{fig5} the behavior of
Eq.~(\ref{chiralintereb}), displayed as the long-dashed curve, resembles the
strong field behavior  for large values of
$eB$.  In order to see the difference between Eqs.~(\ref{chiral21Te}) and
(\ref{chiralintereb}), we consider the regime where ${\tilde m}^2\ll eB$ in
Eq.~(\ref{chiralintereb}), which yields the following leading contribution to the condensate,
\bea
  \Delta\langle\bar{\psi} \psi\rangle_B^T
    &=&-\sqrt{\frac{24}{\pi^3}}\frac{eB}{8\pi}
        \frac{{\tilde m}}{T}+\mathcal{O}\left(eB\frac{{\tilde m}^3}{T^3}\right)\;.
\label{chiralinterec}
\eea
Comparing Eqs.~(\ref{chiralinterec}) and~(\ref{chiralintered}),
we see that
\bea
    \frac{\{\Delta\langle\bar{\psi} \psi\rangle_B^T\}_{inter}}
         {\{\Delta\langle\bar{\psi} \psi\rangle_B^T \}_{strong}}
          =\sqrt{\frac{24}{\pi^3}}\approx 0.879794\;,
\label{chiralinteree}
\eea

\noindent
so that the asymptotic behavior of Eq.~(\ref{chiralintereb})
for large values of $eB$ has an error of $12\%$ compared to strong
field approximation (Eq.~(\ref{chiralintered})) which is the exact
behavior in that region. It may appear that for large magnetic field,
the curve with  intermediate magnetic field strength approximates
the exact result better than the one with large field limit! This is
not quite the case. This is depicted
in Fig.~\ref{fig10}, where we show the relative error,
\be
\varepsilon=\frac{\{\Delta\langle\bar{\psi} \psi\rangle_B^T\}_{exact}-\{\Delta\langle\bar{\psi}
\psi\rangle_B^T\}_{asymp}}{\{\Delta\langle\bar{\psi} \psi\rangle_B^T\}_{exact}}\;,
\ee
of the condensate as a function of the field strength for the asymptotic expressions
of strong and intermediate field regions, Eqs.~(\ref{chiralintered}) and~(\ref{chiralinterec}),
respectively. The curve with the intermediate field strength in fact overshoots the
exact result for large magnetic fields.

\subsection{Weak field limit}

In the weak field limit, we perform a Taylor's
series for $eB\ll {\tilde m}^2$ in Eq.~(\ref{condensate21Tebpoiss}) and we get
\bea
\Delta\langle\bar{\psi} \psi\rangle_B^T&=&
\Delta\langle\bar{\psi} \psi\rangle^T\nonumber \\
     &+&\frac{(eB)^2\tilde{n}({\tilde m})}{12 \pi {\tilde m}^2}
     \left[1+\frac{{\tilde m}}{T}(1-\tilde{n}({\tilde m}))\right]\nonumber \\
     &-&\frac{1}{24 \pi}\frac{(eB)^2}{{\tilde m}^2}\;.
\label{chiral21Tf}
\eea
This expression is shown as a dotted line in Fig.~(\ref{fig5}), which neatly matches onto the exact result
for weak fields.

\end{document}